\documentclass{ws-ijmpd}
\usepackage[super,compress]{cite}

\usepackage{graphicx}
\usepackage{mathtools, amssymb, amsmath}
\usepackage[colorlinks=true, linkcolor=black, urlcolor=blue, citecolor=blue]{hyperref}

\newcommand{\overbar}[1]{\mkern 1.5mu\overline{\mkern-1.5mu#1\mkern-1.5mu}\mkern 1.5mu}
\newcommand{\eref}[1]{Equation~\eqref{#1}} 
\newcommand{\Eref}[1]{Equation~\eqref{#1}} 
\newcommand{\sref}[1]{Section~\ref{#1}} 
\newcommand{\fref}[1]{Fig.~\ref{#1}} 
\newcommand{\tref}[1]{Table~\ref{#1}} 
\newcommand{\Tref}[1]{Table~\ref{#1}} 

\newcommand{\Cref}[1]{Ref.~\refcite{#1}} 

\newcommand{\oG}{\overbar{g}}

\newcommand{\ethree}{\varepsilon^\text{3D}}
\newcommand{\e}{\varepsilon}

\newcommand{\oR}{\overbar{r}}
\newcommand{\oW}{\overbar{w}}
\newcommand{\oL}{\overbar{l}}
\newcommand{\oGamma}{\overbar{\Gamma}}

\numberwithin{equation}{section}

\begin{document}
	
\markboth{T G Tenev, M F Horstemeyer}
{Dark Matter Effect}

%
\catchline{}{}{}{}{}
%

\title{Dark Matter Effect Attributed to the Inherent Structure of Cosmic Space}
	
\author{T G Tenev\footnote{Mississippi State University, Starkville, MS 39759, USA}}

\address{Mississippi State University\\
	Starkville, MS 39759,
	USA \\
	ticho@tenev.com}

\author{M F Horstemeyer}

\address{Mississippi State University\\
	Starkville, MS 39759, USA \\
	mfhorst@me.msstate.edu}

\maketitle

\begin{history}
	\received{Day Month Year}
	\revised{Day Month Year}
\end{history}

\begin{abstract}
We propose that anomalous gravitational effects currently attributed to dark matter can alternatively be explained as a manifestation of the inherent structure of space at galactic length scales. Specifically, we show that the inherent curvature of space amplifies the gravity of ordinary matter such that the effect resembles the presence of the hypothetical hidden mass. Our study is conducted in the context of weak gravity, nearly static conditions, and spherically symmetric configuration, and leverages the Cosmic Fabric model of space developed by Tenev and Horstemeyer [T. G. Tenev and M. F. Horstemeyer, {\it Int. J. Mod. Phys. D} {\bf 27} (2018) 1850083; T. G. Tenev and M. F. Horstemeyer, {\it Rep. Adv. Phys. Sci.} {\bf 2} (2018) 1850011]
	\ccode{PACS numbers:
		04.50.Kd; 
		95.35.+d} 
	
\end{abstract}

\maketitle

\section{Introduction}\label{sec:introduction}

De Swart et al.~\cite{DeSwart2017} present a good review of the historical analysis of dark matter. The notion of dark matter (DM) was introduced in the 1920's and 1930's by Kapteyn~\cite{Kapteyn1922}, Oort~\cite{oort1927observational, Oort1932}, Zwicky~\cite{Zwicky1933, Zwicky1937}, Holmberg~\cite{holmberg1937study}, and Smith~\cite{smith1936no}. Later, in the 1970's and 1980's, it was popularized by Rubin and Ford\cite{Rubin1970, Rubin1980} as a way to explain anomalous rotational curves of galaxies. In addition, DM  was also invoked~\cite{Liddle2015} to explain gravitational lensing, which was discovered by Lynds and Petrosian~\cite{LyndsRandPetrosian1986}. Herein, we will use the term ``Dark Matter effect'' (DM effect) to describe such observations of anomalous gravity. The development of the Standard Cosmological Model, also known as $\Lambda$CDM (where ``CDM'' stands for Cold Dark Matter), appealed to DM as a means to balance the contents of the universe and provide a mechanism for structure formation during the early epoch since its inception~\cite{Liddle2015}. In this context, the $\Lambda$CDM model predicted that DM must be non-baryonic and has to make up a certain fraction of the total contents of the cosmos, namely about 27\%, which is more than 5 times the ordinary (baryonic) matter which, according to the $\Lambda$CDM model, is supposed to comprise only about 5\% of the cosmic content~\cite{Liddle2015}. Whereas the DM effect provides an observational support for DM, by contrast, the idea that DM must make up 27\% of the contents of the universe is model dependent. 

Despite the overwhelming evidence for the DM effect, there has been no direct confirmation~\cite{Liu2017} for the existence of DM, such as would be, for example, the discovery of the particle responsible for DM. The lack of direct evidence for DM has prompted the development of other models to explain the DM effect, such as the Modified Newtonian Dynamics (MOND) theory introduced in the 1980's by Milgrom~\cite{Milgrom1983a}. According to MOND, Newton's Second Law of motion must be modified so that in the case of very weak acceleration $a$ such that $a \ll a_0$, where $a_0$ is a universal acceleration scale parameter, the force $F$ associated with $a$ is no longer linear with respect to $a$ but is proportional to its square. A more narrow formulation of MOND is one where only Newton's Gravitational Law needs to be modified as follows:
\begin{equation}\label{eq:mond-gravity}
ma = F = \frac{G M m}{\mu(a/a_0) \oL^2}, \quad a_0 = 1.2 \times 10^{-10} \text{m}\,\text{s}^{-2}
\end{equation}
where $M$ and $m$ are, respectively, the mass of a gravitating body and that of a test particle situated a distance $\oL$ from each other and attracted to each other with force $F$. Also, $\mu(x)$ is an interpolation function such that  $\mu(x) \to x$ for $x \ll 0$, and $\mu(x)\to 1$ otherwise. Using this simple modification to the Gravity Law, MOND has been successful in explaining the dark matter effect for a great majority of observations~\cite{Milgrom2014}. However, there have also been notable outliers, such as galaxies appearing to have too little~\cite{VanDokkum2018} or too much~\cite{MohammedAli2018} dark matter to fit into MOND's simple one-parameter model. Randriamampandry and Carignan~\cite{Randriamampandry2014} show that among a sample of fifteen galaxies, six do not fit well MOND if $a_0$ were treated as a universal constant but prefer larger or smaller values for it compared to the one given in \eref{eq:mond-gravity}. Problems such as these, and also MOND's empirical nature, that is the lack of satisfactory explanation from first principles, have been continual sources of criticisms. Most recently, Boran et al.~\cite{Boran2018} have argued that the detection of gravitational waves known as GW170817~\cite{Abbott2017} has falsified MOND~\cite{Milgrom1983a} and other~\cite{Bekenstein2004, Moffat2006} ``Dark Matter Emulator'' theories as they call them, because these would have predicted, contrary to observations, that photons and gravitational waves move along different geodesics.

Herein we propose an alternative explanation to the DM effect, which we call the ``Inherent Structure Hypothesis'' (ISH). The ISH is the idea that physical space has inherent structure, such as inherent curvature, that exists apart from matter and leads to modified gravity effects. While the ISH refers to inherent structure in general, herein we focus specifically on inherent curvature being one of its quantifiable attributes.

The Cosmic Fabric analogy of General Relativity proposed by Tenev and Horstemeyer~\cite{Tenev2018, Tenev2018a} helps motivate and analyze the effect of the Inherent Structure Hypothesis. It is a formal analogy interpreting space as a solid body, and the field equations of General Relativity as the bending equation governing the dynamics of said body. In this context, if solid matter can have structure, and space is like a solid object, then it is reasonable to suppose that space too has structure. The work presented here fits within the current limitations of the Cosmic Fabric model~\cite{Tenev2018, Tenev2018a}, namely weak gravity and slow velocities, because the dark matter effect is observed at such conditions. 
The inherent curvature of space represents a modification to the field equations of General Relativity (GR), because it implies that in the absence of any matter-energy fields, the components of the Ricci curvature tensor do not all vanish, but represent the inherent curvature of space. Since the Cosmic Fabric model is an analogy of GR, any results derived through it should also be derivable from conventional GR once its field equations have been modified to account for background spatial curvature. Because the ISH does not invoke new physics, but only new initial configuration, namely an initially curved physical space, it therefore avoids the flaw that Boran et al.~\cite{Boran2018} point out regarding other DM emulator theories.

The notion of inherent structure must be clarified in the context of the notion of length scale, because there can be diverse kinds of structures depending on the length scale. By ``length scale'' we understand a specific range of distances for which certain physical parameters and laws dominate, while others are of lesser significance. For our purpose, we consider the following four length-scales: substructure ($10^{-36}\text{m} - 10^{-10}\text{m}$), continuum ($10^{-10}\text{m} - 10^{14} \text{m}$) , structure ($10^{14}\text{m}-3\times 10^{24}\text{m} $), and cosmic ($3 \times 10^{24}\text{m} - 10^{27} \text{m} $) length-scales. The specific ranges are indicated for the sake of concreteness, but are not intended to be precise. By analogy, the substructure length scale in a conventional material corresponds to the discrete entities comprising the material. Tenev and Horstemeyer~\cite{Tenev2018a} discuss briefly the ramification of physical space having substructure. A more extensive treatment is a subject of subatomic physics and is beyond the scope of this paper.  At continuum length scale, as the name suggests, physical space is treated as a differentiable manifold. General Relativity is strictly a continuum scale theory, and at this length scale, the Cosmic Fabric model~\cite{Tenev2018} yields equivalent results with it. The structure length scale in a conventional material describes the components of which a mechanical system is built, such as the trusses in a bridge, for example. The behavior of these components depends not only on the continuum properties of their material but also on their shape. Our investigation of the Inherent Structure Hypothesis focuses on this length scale, where we have supposed that the space medium forms certain structures whose intrinsic curvatures can be measured and which in fact manifest as the effects currently attributed to dark matter. Finally, the cosmic length scale pertains to the global geometry of the cosmos. To use an analogy: the relationship between the global geometry of the cosmos versus the geometry at its structure length scale is like the relationship between the Earth's global geometry, which is approximately spherical, versus that of the local terrain at various regions on the Earth's surface.

The idea that the DM effect has a geometrical explanation is not new, but has received relatively little attention so far. For example, Bohmer et al.~\cite{Bohmer2008} and later Usman~\cite{Usman2016} propose an ``$f(R)$ modified theory of gravity'' to explain the DM effect where the Ricci scalar $R$, which figures in the Einstein-Hilbert action, is replaced with some more general expression $f(R)$. In a limited sense, our approach can be viewed as a special case of an $f(R)$ theory provided that the inherent curvature of space were constant and can be incorporated as a parameter into $f(R)$. However, per the ISH proposed here, the inherent curvature must be a field, and so the ISH is not the same as an $f(R)$ theory. Dolginov~\cite{Dolginov2011} does consider the inherent geometry of space as the cause for the DM effect and offers several arguments against the conventional DM explanation, such as the absence of dense dark matter clouds. He states that such problems do not exist if the ``the dark matter effect is a result of local non-flat geometry of the empty space.'' However, Dolginov~\cite{Dolginov2011} goes only as far as to raise the possibility for the role that inherent structure plays, but comes short of quantifying the effect, and does not compare it to existing DM models as we have done here.

This paper presents the case for the Inherent Structure Hypothesis (ISH), namely that the Dark Matter (DM) effect is the manifestation of the inherent structure of cosmic space above continuum length scale. We show that the inherent curvature of space amplifies the gravity of ordinary matter that resides within it and we quantify the effect demonstrating that it can feasibly be one and the same as the DM effect. By ``inherent curvature'' we mean the curvature of space that is uncaused by any matter inclusions. In the context of the Cosmic Fabric analogy of physical space, the inherent curvature corresponds to the neutral shape of the cosmic medium prior to it being tensed or compressed. As part of our presentation, we analyze the range of observations for which the ISH produces equivalent results to other models, namely the DM and MOND explanations, and we propose ways in which the ISH can be experimentally distinguished from these models. In order that we can work with closed form expressions, our calculations are for a spherically symmetric configuration and nearly static conditions, but such limitations are not fundamental to the ideas presented here. Furthermore, although we develop the Inherent Structure Hypothesis in the context of the Cosmic Fabric model, its validity does not depend on said model and the same conclusions can be reached by considering solutions to the GR equations that have been modified to account for inherent intrinsic curvature. 

In the remainder of the paper, we introduce the mathematical tools used for handling spherically symmetric inherent curvature (\sref{sec:math}), after which we derive the expressions for how said curvature affects gravity of ordinary matter (\sref{sec:gravity}) and how much of it is required to reproduce the effect of a given hypothetical DM distribution. In the Discussion section (\sref{sec:discussion}) we analyze the conditions under which the Inherent Structure Hypothesis (ISH) proposed here is observationally equivalent with DM and we offer ways to distinguish between the two explanations; we also compare ISH with MOND in the context of sample galactic data, and discuss implication to cosmological models and future work. Finally, we summarize and conclude in \sref{sec:summary}.

\section{Spherically symmetric inherent curvature}\label{sec:math}

\begin{figure}[t]
	\centering
	\includegraphics[width=.6\linewidth]{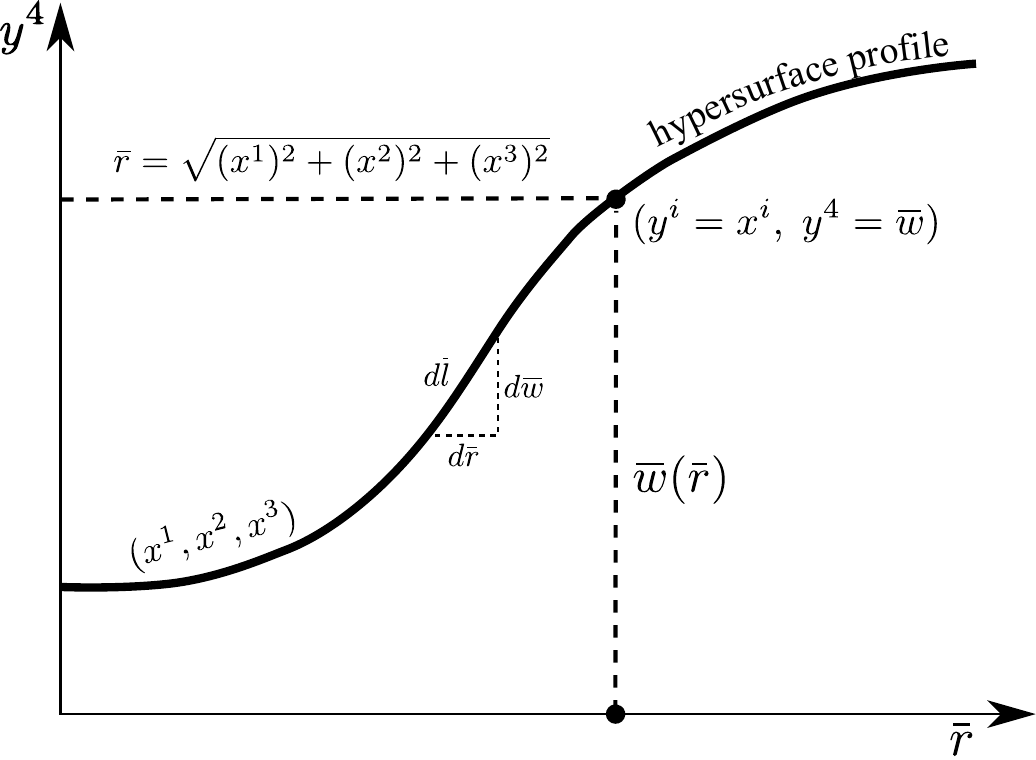}
	\caption{Profile of a spherically symmetric hypersurface (radially symmetric in four-dimensional space) with material coordinates $(x^1, x^2, x^3)$ immersed within a four dimensional reference space with coordinates $(y^1, y^2, y^3, y^4)$. The $y^4$ axis has been aligned with the axis of symmetry and $y^4 = w(\oR)$, where $\oR$ is the distance from the symmetry axis. The hypersurface coordinates $x^i$ have been assigned such that $x^i = y^i$. The relationship between the radial distance element $d\oR$, transverse displacement element $d\oW$, and proper length element $d\oL$ is also indicated; if $d\oL^2 = d\oW^2 + d\oR^2$, then $\oL' = 1 + \oW'$ where the apostrophe indicates differentiation with respect to $\oR$. }
	\label{fig:profile}
\end{figure}

We adapt the coordinate conventions described in~\cite{Tenev2018} for a cosmic fabric with a spherically symmetric inherent curvature. The cosmic fabric (physical space) is considered as immersed in a four dimensional hyperspace within which it can deform. The enclosing hyperspace is flat and has been assigned Cartesian coordinates $y^K,\,K=1\ldots 4$. Within this space, the fabric's spherical symmetry manifests as radial symmetry whose profile is visualized in \fref{fig:profile}. Let $y^K$ be such that $y^4$ is aligned with the axis of symmetry and $y^4 = \oW(\oR)$, where $\oR$ is the distance from $y^4$. Another set of coordinates $x^i,\,i=1\ldots 3$ is painted on the fabric, such that $x^i = y^i$. The time coordinate of the fabric, $x^0$, is defined as usual such that $x^0 \equiv c t$, where $c$ is the speed of light and $t$ is time.

Here, and for the remainder of the paper we have adopted the following notational convention: A bar over the variable name of a quantity indicates that it pertains to the inherent curvature (undeformed configuration) of the cosmic fabric. Upper-case Latin indexes run over the four dimensions of hyperspace $(1\ldots 4)$, lower-case Latin indexes run over the three ordinary spatial dimensions $(1\ldots 3)$, and Greek indexes run over the four spacetime dimensions $(0\ldots 3)$, where the $0^\text{th}$ dimension is time scaled by the speed of light $c$, so it has units of space. Since the fabric represents physical three-dimensional (3D) space, $x^i$ are the coordinates which we have assigned to 3D space in this manner.

\subsection{Derivatives of Radial Functions}

Spherical symmetry allows us to define and analyze the configuration in terms of radial functions that only depend on the distance $\oR$ from the symmetry center. 

Consider the radial function $f = f(\oR)$. Given that $\oR^2 =  \sum_{i=1}^3 (x^i)^2$, therefore:
\begin{equation}\label{eq:derivatives}
\begin{split}
\partial_i \oR &= \frac{x^i}{\oR} \\
\partial_i f &= f'\partial_i \oR = f' \frac{x^i}{\oR} \\
\partial_{ii} f &= f''\frac{(x^i)^2}{\oR^2} + f'\frac{1}{\oR} - f'\frac{(x^i)^2}{\oR^3},\;\text{(no summation)} \\
\nabla^2 f &= f'' + 3 f'\frac{1}{\oR} - f'\frac{1}{\oR} = f'' + 2f'\frac{1}{\oR}
\end{split}
\end{equation}
Also, due to the spherical symmetry and without loss of generality, we will only need the values of the above derivatives at conveniently chosen coordinates, such as the following:
\begin{equation}\label{eq:coords}
x^1 = \oR; \quad x^2 = x^3 = 0
\end{equation}
At these coordinates, the derivatives of $f$ are as follows:
\begin{equation}
\begin{split}
\partial_1 f = f'; \quad
\partial_{11} f = f''; \quad
\partial_{22} f = \partial_{33} f = \frac{1}{\oR}f'
\end{split}
\end{equation}
where apostrophe indicates differentiation with respect to $\oR$.

\subsection{Derivatives of the Reference Coordinates}

The derivatives of the reference coordinates $y^K$ are used to calculate various geometric quantities. We first evaluate the general form of these derivatives, after which we compute them for the choice of coordinates in \eref{eq:coords}. 

Let $\partial_i y^K \equiv \frac{\partial}{\partial x^i} y^K$ and $\oW' \equiv \frac{d}{d\oR}\oW$, and likewise for the second derivatives. Then, 
\begin{equation}
\begin{split}
\partial_i y^4 & = \frac{x^i}{\oR} \oW'
\\
\partial_{ij} y^4 &= \frac{x^i x^j}{\oR^2}\oW'' + \frac{1}{\oR}\oW'\left(\delta_{ij} - \frac{x^i x^j}{\oR^2}\right)
\\
\partial_i y^k &= \delta_{ki}
\end{split}
\end{equation}
where $\delta_{ij}$ is the Kronecker delta. At the chosen coordinates \eqref{eq:coords}, the derivatives of $y^K$ become as follows:
\begin{equation}
\begin{split}
[\partial_i y^K] &= \left(
\begin{array}{cccc}
1 & 0 & 0 & \oW'r' \\
0 & 1 & 0 & 0 \\
0 & 0 & 1 & 0
\end{array}
\right);
\quad
[\partial_{ij} y^4] = \left(
\begin{array}{ccc}
\oW'' & 0 & 0 \\
0 & \frac{1}{\oR} \oW' & 0 \\
0 & 0 & \frac{1}{\oR} \oW'
\end{array}
\right); \quad \partial_{ij} y^k = 0
\end{split}
\end{equation}

\subsection{Spatial Metric Tensor}
The undeformed spatial metric $\oG_{ij}$ can be computed as the inner product of the three surface tangent vectors $\partial_i y^L$:
\begin{equation}\label{eq:g-from-y}
\oG_{ij} = \partial_i y^L \partial_j y^L
\end{equation}
For the chosen coordinates \eqref{eq:coords}, 
\begin{equation}\label{eq:g_ij}
\begin{split}
[\oG_{ij}] &= \left(
\begin{array}{ccc}
1+(\oW')^2 & 0 & 0 \\
0 & 1 & 0 \\
0 & 0 & 1
\end{array}
\right) = 
\left(
\begin{array}{ccc}
(\oL')^2 & 0 & 0 \\
0 & 1 & 0 \\
0 & 0 & 1
\end{array}
\right)
\\
[\oG^{ij}] &= [\oG_{ij}]^{-1}
= \left(
\begin{array}{ccc}
(\oL')^{-2} & 0 & 0 \\
0 & 1 & 0 \\
0 & 0 & 1
\end{array}
\right)
\end{split}
\end{equation}
where we have made the following substitution:
\begin{equation}\label{eq:l-from-w}
(\oL')^2 = (\oW')^2 + 1
\end{equation}
such that $\oL = \oL(\oR)$ represents the proper radial distance from the center of symmetry, that is, the distance as measured from within the fabric (see \fref{fig:profile}).

The first derivatives of the metric can be computed by differentiating \Eref{eq:g-from-y} and evaluating at the special coordinates~\eqref{eq:coords}. The only non-vanishing derivatives are the following:
\begin{equation}\label{eq:partial_g}
\begin{split}
\partial_1 \oG_{11} &= 2 \oL'\oL'' \\
\partial_2 \oG_{12} &= \partial_2 \oG_{21} = \partial_3 \oG_{13} = \partial_3 \oG_{31} = \frac{(\oL')^2 - 1}{\oR}
\end{split}
\end{equation}

\subsection{Christoffel Symbols}

The Christoffel symbols $\oGamma^m_{\;ij}$ characterize how inherent curvature affects field derivatives. These can be calculated from the metric as follows:
\begin{equation}
\oGamma^m_{\;ij} = \frac{1}{2} \oG^{mk}\left(\partial_j \oG_{ki} + \partial_i \oG_{jk} - \partial_k \oG_{ij}\right),
\end{equation}
We proceed to evaluate these for the special coordinate choice~\eqref{eq:coords}. For the chosen coordinates, the metric is diagonal so we will only need to evaluate those Christoffel symbols for which $i = j$.  We do so using the metric values from Equations \eqref{eq:g_ij} and \eqref{eq:partial_g}. Of the evaluated Christoffel symbols, only the following are non-vanishing:
\begin{equation}\label{eq:christoffel}
\begin{split}
\oGamma^1_{11} &= \frac{1}{2}\oG^{11}\partial_1 \oG_{11} = \frac{\oL'\oL''}{(\oL)^2} \\
\oGamma^1_{22} &= \oG^{11}\partial_2 \oG_{12} = \frac{(\oL')^2-1}{\oR (\oL')^2} \\
\oGamma^1_{33} &= \oG^{11}\partial_3 \oG_{13} = \frac{(\oL')^2-1}{\oR (\oL')^2} \\
\end{split}
\end{equation}

\section{Gravity in the context of inherent spherically symmetric curvature}\label{sec:gravity}

Consider physical space with radially symmetric inherent curvature specified by the displacement function $\oW$ as per the discussion in \sref{sec:math}, which has been deformed due to the presence of a gravitating mass where $g_{ij}$ represents its deformed metric, and $g_{\mu\nu}$ is the metric of the resulting spacetime. A free-falling particle moves along a spacetime geodesic whose equation is given by,
\begin{equation}\label{eq:geodesic}
\begin{split}
& \ddot{x}^\alpha + \Gamma^{\alpha}_{\mu\nu} \dot{x}^\mu \dot{x}^\nu = 0,
\quad
 \text{such that}\quad g_{\alpha\beta}\Gamma^{\alpha}_{\mu\nu} = \frac{1}{2} 
\left( \partial_\nu g_{\mu\beta} + \partial_{\mu} g_{\beta\nu} - \partial_\beta g_{\mu\nu} \right)
\end{split}
\end{equation}
where the dot notation represents differentiation with respect to proper time. Consider a particle initially at rest with respect to the fabric and located at the chosen coordinates~\eqref{eq:coords}. Due to the spherical symmetry, the following reasoning applies to any particle that is a distance $\oR$ from the symmetry center. Because the particle is at rest, its initial four-velocity is such that $\dot{x}^0 = c$ and $\dot{x}^i = 0$, where $c$ is the speed of light. Under these circumstances, \eref{eq:geodesic} reduces to the following,
\begin{equation}\label{eq:coord-accel}
\begin{split}
\ddot{x}^1 +c^2\Gamma^1_{00} &= 0, 
\quad  \text{s.t.} \quad g_{11}\Gamma^{1}_{00} = \frac{1}{2}\left(- \partial_1 g_{00} \right) \\
\therefore \ddot{x}^1 &= c^2 \frac{\partial_1 g_{00}}{2g_{11}} \approx c^2 \frac{\partial_1 g_{00}}{2 \oG_{11}} = c^2 \frac{\partial_1 g_{00}}{2 (\oL')^2} 
\end{split}
\end{equation}
The approximation in \eqref{eq:coord-accel} invokes the weak gravity (small strains) assumption due to which the deformed and undeformed metrics are nearly identical, $g_{ij} \approx \oG_{ij}$. However, note that such approximation does not necessarily apply for the spatial derivatives of $g_{ij}$ and $\oG_{ij}$.

Let $a \equiv \ddot{l}$ be the proper radial acceleration, where $l$ stands for the deformed radial distance. Again, due to the weak gravity (small strains) assumption, we can approximate $l \approx \oL$ (the deformed and undeformed proper distances are about the same), and because of the assumed nearly static conditions, the approximation can be carried to the time derivatives so that $\ddot{l} \approx \ddot{\oL}$. Under the nearly static conditions, $\oL'$ also does not change significantly in time, so $\ddot{\oL} \approx \ddot{x}^1 \oL'$ and thus $a \approx \ddot{x}^1 \oL'$, which combined with \eref{eq:coord-accel} produces the following:
\begin{equation}\label{eq:a-from-g_00}
a \approx  \ddot{x}^1 \oL' = c^2 \frac{\partial_1 g_{00}}{2\oL'} = c^2 \frac{\partial_r g_{00}}{2\oL'}
\end{equation}
The last equality takes advantage of the spherical symmetry to generalize the expression for $a$ to any location that is a distance $\oR$ from the symmetry center.

From the Time Lapse postulate of the Cosmic Fabric model~\cite{Tenev2018} one can derive the following relationship between the time-time component of the deformed spacetime metric and the three-dimensional volumetric strain $\e$ (see Equations (2.10) and (2.11) in \Cref{Tenev2018} where $\ethree$ was used to denote the three dimensional volumetric strain, but here we have dropped the superscript ${}^\text{3D}$ for clarity):
\begin{equation}\label{eq:g_00}
g_{00} = -(1 + \e)^{-2} \approx -1 + 2\e
\end{equation}
Recall that the volumetric strain $\e$ is a scalar field characterizing the deformation of the fabric in terms of its fractional volumetric increase at a given location in space. \Eref{eq:g_00} combined with \eref{eq:a-from-g_00} yields the following:
\begin{equation}\label{eq:a-from-ethree}
a = c^2\frac{\e'}{\oL'},
\end{equation}
where we have replaced the approximation sign with an equality sign that applies in the regime of weak gravity (small strains) and nearly static conditions. 
Notice that the radial derivative of the acceleration, $a'$, is as follows:
\begin{equation}\label{eq:a-prime}
a' = c^2\frac{1}{\oL'}\e'' - c^2\frac{\oL''}{(\oL')^2}\e'
\end{equation}

The Inclusion Postulate of the Cosmic Fabric model~\cite{Tenev2018} relates the volumetric strain $\e$ to the density of the gravitating mass. For inherently curved space we must use covariant derivatives, in terms of which the Inclusion Postulate is as follows:
\begin{equation}\label{eq:cov-laplace-ethree}
\nabla_i (\nabla^i \e) = -\frac{1}{2}c^2 \kappa \rho
\end{equation}
where $\nabla_i$ is the covariant derivative with respect to the $x^i$ coordinate, $c$ is the speed of light, $\rho$ is the density of the inclusion, and $\kappa$ is the Einstein constant. The covariant Laplacian can be evaluated from the following identity and using the Christoffel symbols~\eqref{eq:christoffel}:
\begin{equation}\label{eq:cov-laplacian}
\begin{split}
\nabla_i (\nabla^i \e) = & \oG^{ij}\left( \partial_{ij} \e - \oGamma^m_{\;ij} \partial_m \e \right) \\
 = & \oG^{11}\partial_{11}\e - \oG^{11}\oGamma^{1}_{\;11}\partial_1 \e \\
& + \oG^{22}\partial_{22} \e - \oG^{22}\oGamma^1_{\;22}\partial_1 \e \\
& + \oG^{33}\partial_{33} \e - \oG^{33}\oGamma^1_{\;33}\partial_1 \e \\
= & \frac{1}{(\oL')^2}\left[\e'' - \frac{\oL''}{\oL'}\e' + \frac{2}{\oR}\e'\right]
\end{split}
\end{equation}
In the last step of the above derivation, we have used the result from \eref{eq:derivatives} for the derivatives of a radial function. Combining Equations \eqref{eq:a-from-ethree} - \eqref{eq:cov-laplacian}, we arrive at the following surprisingly simple differential equation in terms of the proper acceleration:
\begin{equation}\label{eq:a-ode}
\begin{split}
a' + \frac{2}{\oR}a = -\frac{1}{2}c^4 \kappa\rho \oL'
\end{split}
\end{equation}
whose general solution has the following form:
\begin{equation}\label{eq:a-general-solution}
a(\oR) = -\frac{1}{\oR^2}\left(C_1 + \frac{1}{2}c^4 \kappa \int_0^{\oR} \rho \oL'  \xi^2 d\xi\right)
\end{equation}
where $C_1$ is a constant of integration, and $\xi$ is the integration variable representing the coordinate distance from the center of symmetry. To avoid instability as $\oR \to 0$, we require that $C_1 = 0$. Furthermore, since $\kappa \equiv 8\pi G /c^4$, where $G$ is the gravitational constant, \eref{eq:a-general-solution} becomes the following,
\begin{equation}\label{eq:a-solution}
a(\oR) = -\frac{G}{\oR^2} \int_0^{\oR} \rho (4\pi \xi^2) \oL' d\xi = -\frac{G M(\oR)}{\oR^2}
\end{equation}
where $M(\oR)$ represents the gravitating mass enclosed within the coordinate radius $\oR$. That is because the expression $(4\pi \xi^2) \oL' d\xi$ represents the volume of a spherical shell with coordinate radius $\oR$, surface area $4\pi \xi^2$ and thickness $\oL' d\xi$.

\begin{figure}[t]
	\centering
	\includegraphics[width=0.7\linewidth]{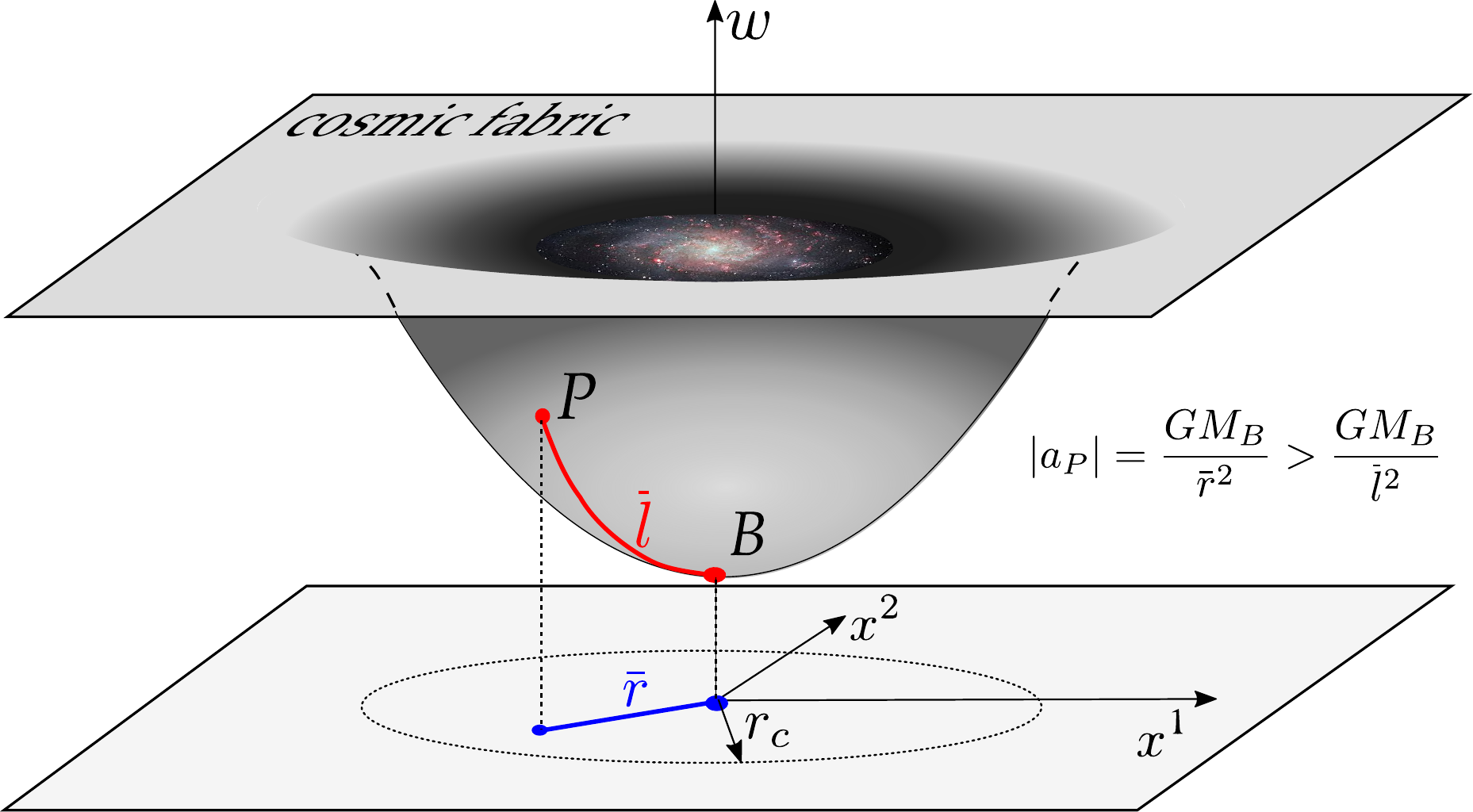}
	\caption{The inherent shape of space causing the ``Dark Matter'' effect. A test particle at $P$ and a proper distances $\oL$ from a body of mass $M_B$ that is located at $B$ experiences gravitational acceleration $a_P$. The magnitude of $a_P$ is greater than predicted by the Inverse Square Law in terms of $\oL$, but matches the predication of said law when the projected (coordinate) distance $\oR$ is used instead of $\oL$. The critical coordinate distance $r_c$ represents the limit within which the Inherent Structure Hypothesis is observationally equivalent to the hypothetical presence of dark matter. Typically, $r_c$ extends beyond the edge of visible galactic matter.}
	\label{fig:dark-matter-effect}
\end{figure}

\Eref{eq:a-solution} is none other than Newton's Gravitational Law but expressed in terms of the coordinate distance to the gravitating mass instead of the proper distance. This result is true for our particular choice of coordinates where the coordinate distance $\oR$ is actually the projection of the proper distance $\oL$ onto a flat hypersurface that is perpendicular to the axis of symmetry, as illustrated in \fref{fig:dark-matter-effect}.

\Eref{eq:a-solution} and \fref{fig:dark-matter-effect} show how the presence of inherent curvature amplifies the gravitational effective of ordinary matter: since $\oR \le \oL$, the resulting acceleration is greater than what one would have expected from applying Newton's Gravity Law with the proper distance $\oL$. The additional  gravitational acceleration might be interpreted as caused by invisible mass, that is dark matter. However, as we demonstrated above, such extra gravitational acceleration can also be explained by the inherent curvature of space.

\section{Discussion}\label{sec:discussion}

Below we compare the Inherent Structure Hypothesis (ISH) to other explanations of the Dark Matter effect, and propose how these can be experimentally distinguished from one another. We also discuss the implication of the ISH to cosmological models. 

\subsection{Conditions for observational equivalence and falsifiability}

We analyze below under what conditions is the Inherent Structure Hypothesis observationally equivalent to the Dark Matter explanation of the Dark Matter effect. For this purpose, we consider the following two questions: 
\begin{enumerate}
	\item Can the effect of any given dark matter distribution be equivalently explained by a geometrically consistent inherent curvature, and
	
	\item Can the effect of any given inherent curvature be equivalently explained by a physically admissible dark matter distribution?
	
\end{enumerate}
By ``geometrically consistent'' we mean that the fabric's material does not intersect itself and has no kinks. Also, for the dark matter distribution to be ``physically admissible,'' we require that it has finite non-negative density. Below, we answer the above questions in the context of spherical symmetry, but we expect that the responses also apply more generally.

To answer the first question above, consider the radial functions $M_\text{DM} = M_\text{DM}(\oR)$ and $M = M(\oR)$ representing, respectively, the hypothetical dark matter mass and ordinary mass enclosed within some coordinate radius $\oR$ that corresponds to proper distance from the center $\oL = \oL(\oR)$. According to the Dark Matter Hypothesis, one would expect that the proper acceleration is $a = -G(M + M_\text{DM})/\oL^2$, which in view of \eref{eq:a-solution} implies the following:
\begin{equation}\label{eq:l-from-r}
\oL = \oR \sqrt{1 + \frac{M_\text{DM}}{M}}
\end{equation}
\begin{equation}\label{eq:lprime-from-r}
\oL' = \sqrt{1 + \frac{M_\text{DM}}{M}} + \frac{\oR(M_\text{DM}/M)'}{2\sqrt{1 + M_\text{DM}/M}}
\end{equation}
\Eref{eq:lprime-from-r} fully specifies the equivalent inherent curvature, because once $\oL'$ is known, the displacement function $\oW$ can be computed from \eref{eq:l-from-w} up to a rigid translation. For the resulting curvature to be geometrically consistent, we require that $1 \le \oL' < \infty$. From \eref{eq:lprime-from-r}, it is clear that except in the complete absence of visible matter ($M  = 0$) it must be the case that $\oL' < \infty$. Furthermore, as long as the ratio between dark to visible matter increases with distance from the center, then $(M_\text{DM}/M)'\ge 0$ and so $\oL' \ge 1$. Both of these are consistent with known observations, because the Dark Matter effect is always observed along with visible matter. Also, dark matter is supposed to dominate the exterior of the galaxies, so that the ratio of the dark to visible matter content enclosed within a given radius increases in the outward direction. Therefore, any known distribution of dark matter can be modeled as inherent curvature.

To answer the second question above, we now consider how an inherent curvature profile specified by the proper distance function $\oL = \oL(\oR)$ is interpreted as dark matter content $M_\text{DM}(\oR)$. For simplicity, we will focus on distances sufficiently far away from the center of symmetry beyond which the enclosed visible mass does not increase appreciably so that $M(\oR) \approx const$. This simplification is consistent with the structure of galaxies and dark matter models where most of the visible mass is concentrated within the galactic center, while the dark matter halo is supposed to extend well beyond the visible mass of the galaxy. By rearranging \eref{eq:l-from-r} we obtain the following:
\begin{equation}
\begin{split}
M_\text{DM} = M\left(\frac{\oL^2}{\oR^2} - 1\right); \quad
M'_\text{DM} = 2\frac{M l}{\oR^2}\left(\oL' - \frac{l}{\oR}\right)
\end{split}
\end{equation}
For $M_\text{DM}$ to be physically admissible, we require that  $M'_\text{DM} \ge 0$, since the reverse implies negative dark matter density. Therefore, we require that,
\begin{equation}\label{eq:lprime-condition}
\oL' \ge \frac{l}{\oR}
\end{equation}
Beyond the boundary of the hypothetical dark matter halo, $M'_\text{DM} = 0$, so $\oL' = \oL/\oR$ implying that $\oL' = const$. Therefore, within the dark matter halo where $M'_\text{DM} > 0$, we would expect that $\oL'$ is monotonically increasing. So, in general, $\oL'$ has to be non-decreasing for the DM effect due to inherent curvature to be explainable by actual dark matter. In other words, we conclude the following:
\begin{equation}\label{eq:lsecond-condition}
\oL'' \ge 0
\end{equation}
Given the relationship between $\oW$ and $\oL$ where $(\oW')^2 = (\oL')^2 - 1$, \eref{eq:lsecond-condition} means that the radial function $\oW(\oR)$ should not change concavity for the equivalent dark matter to be physically admissible as such. 

Unless the inherent curvature of space can be maintained globally for the entire cosmos, it will necessarily be the case that beyond certain critical radius $r_c$ the condition stated in \eref{eq:lsecond-condition} no longer applies. As \fref{fig:dark-matter-effect} and \fref{fig:mond-profile} illustrate, beyond $r_c$, the concavity of $\oW(\oR)$ must reverse for the local inherent curvature  to return to flat or to some lesser curvature that can be maintained globally. Therefore, beyond such $r_c$ the Inherent Structure and the Dark Matter hypotheses are no longer observationally equivalent. 

Beyond the critical radius $r_c$, the expected observational differences between the Inherent Structure and the Dark Matter explanations, provide a way to verify one and falsify the other. In particular, per the Inherent Structure Hypothesis (ISH):
\begin{enumerate}
	\item At sufficiently large distances from the center of the hypothetical dark matter halo, the dark matter effect will reverse and eventually disappear as if the halo were not present. In other words, a test particle orbiting a galaxy well beyond the critical distance $r_c$ will behave as if no dark matter were enclosed within its orbit.
	
	\item Gravitational systems for which the DM effect is observed will exhibit a relatively more pronounced edge at approximately the critical distance $r_c$ compared to gravitational systems that do not exhibiting the DM effect.
	
	\item On the cosmic length-scale, the hypothetical ``dark matter'' will have no net contributions to the contents of the universe. Note that the currently estimated 27\% dark matter content is a model-dependent result and therefore does not necessarily falsify the ISH. 
	
\end{enumerate}

Confirming or falsifying any of the above predictions will either confirm or falsify the Inherent Structure hypothesis.

\subsection{Comparison with Modified Newtonian Dynamics (MOND)}

\begin{figure}[t]
	\centering
	\includegraphics[width=0.6\linewidth]{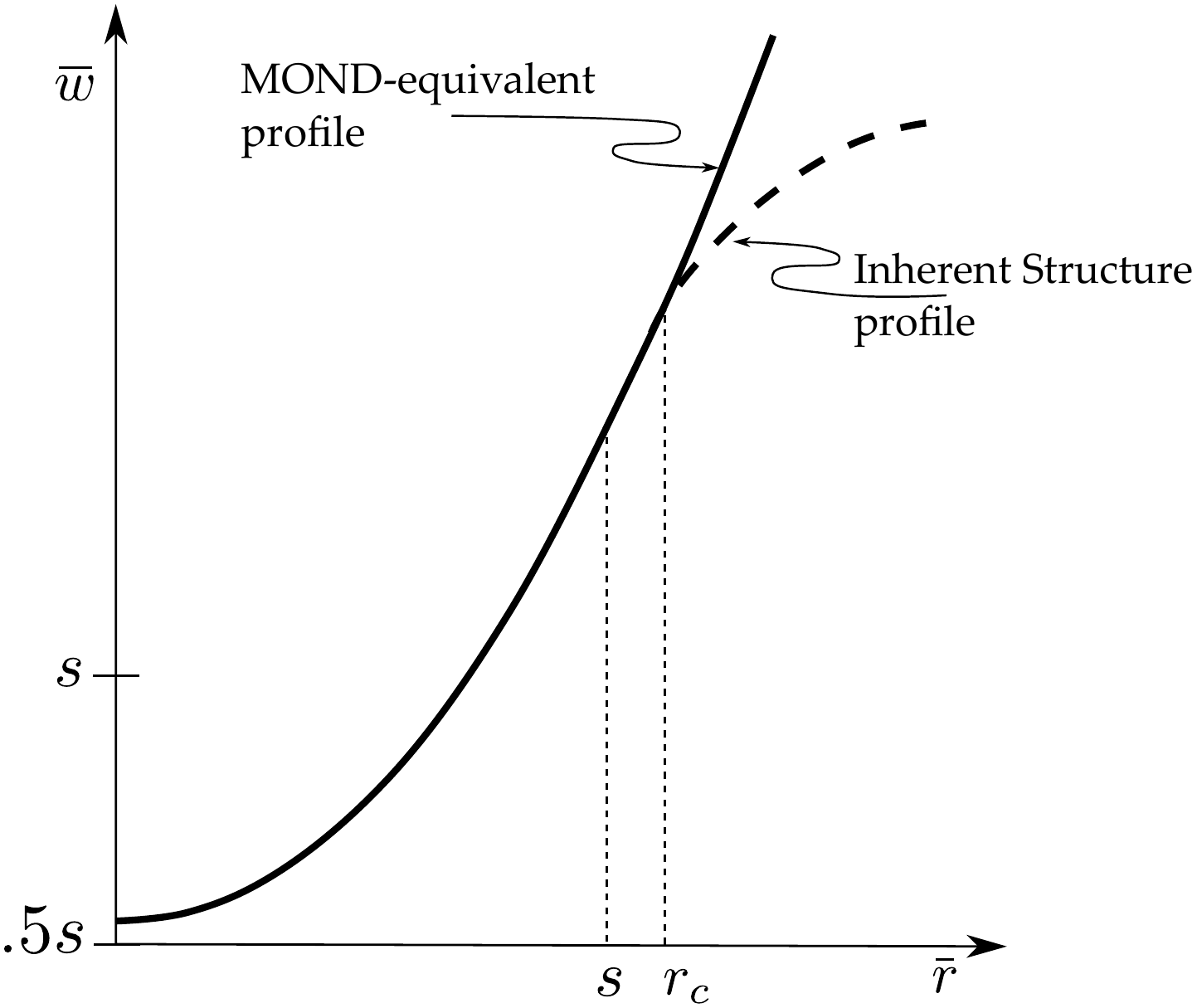}
	\caption{The inherent curvature profile whose effect is equivalent to the MOND model up to a certain critical distance $r_c$. The deviation of the actual inherent structure profile from the MOND-equivalent one takes place beyond the critical distance. The vertical displacement is given by $\oW = \frac{1}{2}s\sqrt{AB} - \frac{1}{4}s\ln\left( 2\sqrt{A} + \sqrt{B} \right)$, where $A \equiv (\oR/s)^2 + 1$, $B \equiv 4(\oR/s)^2 + 3$, $s \equiv \sqrt{GM / a_0}$, and $a_0$, $G$, and $M$ are, respectively, the MOND parameter, the gravitational constant, and the gravitating mass whose gravity is being amplified by the inherent curvature.  }
	\label{fig:mond-profile}
\end{figure}

We show below that the Inherent Structure Hypothesis yields equivalent results with MOND to within a critical radius $r_c$ for an appropriately chosen inherent curvature profile $\oW(\oR)$. For this purpose, we use the following interpolation function:
\begin{equation}\label{eq:interp}
\mu(a/a_0) = \frac{1}{1 + a_0/a}
\end{equation}
Substituting \eref{eq:interp} into \eref{eq:mond-gravity} and solving tor $a$ yields the following:
\begin{equation}\label{eq:MOND-a}
a = -\frac{GM}{2\oL^2}\left(1 + \sqrt{1 + \frac{4a_0 \oL^2}{GM}}\right)
\end{equation}
At the same time, since $a = -GM/\oR^2$ per \eref{eq:a-solution}, where $\oR$ is the coordinate distance, therefore:
\begin{equation}\label{eq:mond-l-vs-r}
\begin{split}
\frac{1}{\oR^2} & = \frac{1}{2\oL^2}\left(1 + \sqrt{1 +\frac{4a_0 \oL^2}{GM}}\right),
\end{split}
\end{equation}
which when solved for $\oL$, produces the following result:
\begin{equation}\label{eq:mond-l-from-r}
\oL = \oR\sqrt{1 + a_0\frac{\oR^2}{G M}}
\end{equation}
From \eref{eq:mond-l-from-r} we can determine $\oL'$, which substituted into \eref{eq:l-from-w} brings us to the following result for the displacement $\oW$:
\begin{equation}
\begin{split}
(\oW')^2 & = \oR^2 \frac{3s^2 + 4 \oR^2}{s (s^2 + \oR^2)}; \quad s \equiv \sqrt{\frac{G M}{a_0}}\\
\oW' &= \frac{\oR}{s}\sqrt{\frac{3 s^2 + 4 \oR^2}{s^2 + \oR^2}}
\end{split}
\end{equation}
where $s$ is a scale parameter characterizing the gravitating mass $M$. The solution for $\oW$, which is also plotted in \fref{fig:mond-profile}, is as follows:
\begin{equation}\label{eq:mond-profile}
\begin{split}
\oW = & \frac{s}{2}\sqrt{AB} - \frac{s}{4}\ln\left( 2\sqrt{A} + \sqrt{B} \right) \\
& \quad A \equiv (\oR/s)^2 + 1; \quad B \equiv 4(\oR/s)^2 + 3,
\end{split}
\end{equation}

The actual profile of the inherent structure of space need not match exactly the MOND-equivalent one; \fref{fig:mond-profile} shows how the two might diverge beyond certain critical distance $r_c$. If $r_c$ is sufficiently far outside the edge of the gravitational system under consideration, then the variance with MOND may not be readily observed. That is why in \fref{fig:mond-profile}, the critical distance is illustrated for the case where $r_c > s$. When the actual inherent structure profile illustrated in \fref{fig:mond-profile} is revolved around the transverse axis, the result is the shape in \fref{fig:dark-matter-effect}.

\Tref{tab-galaxies} shows that within a diverse sample of galaxies, the vast majority conform to the profile illustrated in \fref{fig:mond-profile}. The table shows that for most galaxies, the characteristic scale $s$ is greater than the radius $R$ of the visible galactic mass, and is also within an order of magnitude of it. Note that in this case $R$ represents the proper distance, which is in general longer than the corresponding undeformed coordinate distance. Therefore, $s$ being greater than $R$ also implies that $s$ is greater than the corresponding coordinate distance. In a few cases, like the Cartwheel galaxy, where the galactic radius appears to exceed the characteristic scale $s$, the associated inherent structure of space likely differs from the generic profile in \fref{fig:mond-profile}. In the case of the Cartwheel galaxy, for example, such variance is not surprising because of the unusual shape of that galaxy.

\begin{table}
	\tbl{Apparent masses and sizes of galaxies $25\times 10^3$ - $13.4\times 10^9$ light years from Earth. The list is sorted by distance from Earth. The scale factor $s=\sqrt{GM/a_0}$, where $a_0$ is the MOND parameter, is characteristic of the inherent curvature of space associated with that galaxy. The visible radius $R$ is to within an order of magnitude of the characteristic scale $s$. This relationship appears to hold for a broad variety of galaxies. The acronyms LMC and SMC stand for Large Magellanic Cloud and Small Magellanic Cloud, respectively.}
	{\begin{tabular}{@{}lccc@{}} 
			\toprule
			Galaxy\hphantom{===================} & Mass $M$ & Radius $R$ & $R/s$  \\
			& $[10^9 \text{M}_\odot]$  & $[10^3\;\text{ly}]$ &  \\ 
			\colrule
			Milky Way~\cite{McMillan2011, McMillan2017, Kafle2012, Kafle2014, Nasa2018_MilkyWay} & \hphantom{000}1000 & \hphantom{0}50.0 & \hphantom{0}0.45 \\
			LMC~\cite{PaulW.Hodge2009} & \hphantom{00000}10 & \hphantom{00}7.0 & \hphantom{0}0.63 \\
			SMC~\cite{PaulW.Hodge2009} & \hphantom{000000}7 & \hphantom{00}3.5 & \hphantom{0}0.38 \\
			Andromeda~\cite{Kafle2018a, Penarrubia2014, Chapman2006} & \hphantom{000}1000 & 110.0 & \hphantom{0}0.99 \\
			M33~\cite{Corbelli2003, GerardP.Michon2016} & \hphantom{00000}50 & \hphantom{0}30.0 & \hphantom{0}1.21 \\
			Pinwheel~\cite{Comte1979} & \hphantom{0000}100 & \hphantom{0}85.0 & \hphantom{0}2.42 \\
			Whirlpool~\cite{Nasa_whirlpool, Theplanets.org_whirlpool} & \hphantom{0000}160 & \hphantom{0}30.0 & \hphantom{0}0.67 \\
			Sunflower (M63)~\cite{mobjects2015_sunflower} & \hphantom{0000}140 & \hphantom{0}49.0 & \hphantom{0}1.18 \\
			M77~\cite{mobjects2015_m77} & \hphantom{000}1000 & \hphantom{0}85.0 & \hphantom{0}0.76 \\
			Condor~\cite{Horellou2007, Eufrasio2014} & \hphantom{0000}100 & 261.0 & \hphantom{0}7.42 \\
			Cartwheel~\cite{Amram1998} & \hphantom{000000}4 & \hphantom{0}75.0 & 10.67 \\
			Malin 1~\cite{Barth2007} & \hphantom{000}1000 & 325.0 & \hphantom{0}2.92 \\
			Phoenix Cluster~\cite{McDonald2012} & 2000000 & 550.0 & \hphantom{0}0.11 \\
			GN-z11~\cite{Oesch2016} & \hphantom{000000}1 & \hphantom{00}1.5 & \hphantom{0}0.43 \\
			\botrule
		\end{tabular} \label{tab-galaxies}}
\end{table}

Interestingly, the relationship between $R$ and $s$ illustrated in \tref{tab-galaxies} also holds for our Solar System, which suggests that it may apply to smaller gravitational systems and not just galaxies. For example, in the case of the Solar System, $M \approx 1 \text{M}_\odot$, $R = 1.43\times 10^{14} \text{m}$ (the distance between the Sun and the planet Sedna), and consequently $R/s = 0.14$.

The above comparison between the Inherent Structure Hypothesis (ISH) and MOND serves to validate ISH, because MOND has been empirically shown to provide good explanation for the DM effect in most cases~\cite{Milgrom2014, Milgrom2014a}. 

At the same time, the ISH provides a more general and potentially more precise explanation than MOND. Being a single-parameter model, MOND still leaves a considerable number of outliers~\cite{Randriamampandry2014} where $a_0$ is either too large or too small. By contrast per the ISH such cases simply reflect, respectively, less or more inherent curvature. Furthermore, MOND calls for the modification of a fundamental law of nature, such as Newton's Second Law of motion, or at the very least Newton's Gravity Law, which has far reaching effects. By contrast, the ISH only refers to the properties of a specific object of nature, namely physical space, and only in a specific region. Finally, there are clues within MOND and also from the analysis above to suggest that the likely explanation for the DM effect is geometrical in nature. One such clue is that the MOND parameter $a_0$, which has been empirically derived, when expressed as length, $l_\text{MOND} \equiv c^2/a_0 = 7.5\times 10^{26}\text{m}$, is comparable to the Hubble distance, $l_\text{H} = c/H_0 = 1.4\times 10^{26} \text{m}$ where $H_0$ is the Hubble parameter. At the same time, the Hubble distance is characteristic of the size of the observable universe. Another clue is that the characteristic length scale $s$ computed for each gravitational system based on the MOND hypothesis, happens to be comparable to the geometrical size of said system (see \tref{tab-galaxies}). Both of these ``coincidences'' are empirical as opposed to an artifact of the model, so they point to some geometrical (or structural) characteristics of the underlying reality. MOND's single parameter model seems like a first order approximation for the inherent structure of space.

\subsection{Implication to cosmological models}

Like MOND and other dark matter alternative models, the Inherent Structure Hypothesis (ISH) is  incompatible with the $\Lambda$CDM model, which critically depends on the existence of non-baryonic dark matter~\cite{Liddle2015}. Even more fundamentally, the $\Lambda$CDM model depends on the presupposition known as the Cosmological Principle~\cite{Liddle2015} that at the cosmic length scale (greater than 100 Mpc or $3\times 10^{24}\text{m}$) the universe is homogeneous and isotropic, which in essence is a presupposition about the absence of structure at that scale.

By contrast, ISH is based on the presupposition that structure is a fundamental property of nature, and as such, it is an essential element in cosmological models that adopt the same view. Indeed, common experience shows that every sufficiently complex functioning system exhibits structure at its greatest length scale and so the cosmos should be no exception. For this reason, it is quite likely that new cosmological data will soon conclusively repudiate the Cosmological Principle triggering the revision of the Standard Cosmological Model to account for structure at every length scale. The ISH will be well fitted for such a revised cosmological model.

What would a cosmological model based on the presupposition of structure look like? According the ISH, the inherent curvature of space is uncaused by matter, and yet, the galaxy data in \tref{tab-galaxies} demonstrates a correlation between inherent curvature and matter. Therefore, one must conclude that the there is a causal relationship after all but in reverse: the inherent structure of space is what causes matter to form galaxies and galactic clusters in the first place. This idea may also explain the so called Large Scale Structure of the universe consisting of walls and filaments made up of galaxies and galactic clusters that appear organized into definite forms, but are not gravitationally bound together. New cosmological models may appeal to the inherent structure of space as the seed needed for matter-structure formation similarly to how the $\Lambda$CDM model appeals to dark matter for the same purpose.

\subsection{Future Work}

With the help of numerical simulations, the work presented in this paper can be used to interpret existing observations of the DM effect to create a map of the inherent structure of cosmic space much like echo sonars can create a map of the Earth's ocean floor. For this purpose, the work herein will need to be generalized to non-symmetric configurations, which can be accomplished, for example, by approximating such configurations as the linear superposition of symmetric ones. If the resulting map reveals an inherent cosmic structure that is geometrically consistent, then such discovery will add credence to the Inherent Structure Hypothesis.

Another area of future work is devising experiments that can distinguish between the Inherent Structure Hypothesis (ISH) and other models that attempt to explain the Dark Matter effect. Such experiments would involve, for example, measuring anomalous gravity outside the edges of galaxies to detect whether the supposed inherent curvature begins to attenuate (see the dashed line in \fref{fig:mond-profile}).

\section{Summary and Conclusion}\label{sec:summary}

We showed that the inherent curvature of physical space (that is curvature uncaused by matter) amplifies the gravitational effects of ordinary matter and produces the kind of gravitational anomalies that are currently attributed to the presence of dark matter (DM). We proposed the Inherent Structure Hypothesis (ISH) stating that the so called DM effect is the manifestation of the inherent structure of space at galactic length-scales, and not the result of invisible mass. 

We demonstrated that any DM effect, which can be explained by the Modified Newtonian Dynamics (MOND) theory or by the presence of a DM halo, can be equally well explained by the ISH. At the same time, we showed, ISH allows for DM effects that cannot be explained by MOND or by DM halos. Therefore, we concluded that the Inherent Structure and DM explanations are observationally equivalent with each other to within some distance from the center of a gravitating system. However, beyond such distance, the ISH predicts that the gravitational impact of the hypothetical dark matter begins to be reversed and is nearly completely eliminated at sufficiently far distances. This is a verifiable prediction that would distinguish our model from other explanations of the DM effect.

In the comparison between the ISH and MOND we noted an interesting relationship between the size of a gravitational system and its Schwartzchild radius through the MOND parameter $a_0$. Such relationship hinted at the structural underpinnings of the DM effect. 

The Inherent Structure Hypothesis stems from the principle that structure is a fundamental aspect of matter, space, and nature in general, and as such can be incorporated into cosmological models that subscribe to the same principle.

\bibliographystyle{ws-ijmpd}
\bibliography{dark-matter-effect}

\end{document}